\documentstyle[twoside,fleqn,psfig,espcrc2]{article}%
\begin{document}
\renewcommand{\refname}{\normalsize\bf References}
\title{%
Distribution of the reflection eigenvalues\\
of a weakly absorbing chaotic cavity
}

\author{%
        C.W.J. Beenakker%
        \address{\em Instituut-Lorentz, Universiteit Leiden\\
                 P.O. Box 9506, 2300 RA Leiden, The Netherlands}%
        \,and
        P.W. Brouwer%
        \address{\em Lyman Laboratory of Physics, Harvard University\\
                 Cambridge MA 02138, USA}%
\thanks{Present address: Laboratory of Atomic and Solid State Physics,
        Cornell University, Ithaca NY 14853, USA}
}
%
%
\begin{abstract}
\hrule
\mbox{}\\[-0.2cm]

\noindent{\bf Abstract}\\

The scattering-matrix product $SS^{\dagger}$ of a weakly absorbing medium
is related by a unitary transformation to the time-delay matrix without
absorption. It follows from this relationship that the eigenvalues of
$SS^{\dagger}$ for a weakly absorbing chaotic cavity are distributed
according to a generalized Laguerre ensemble.\\[0.2cm]
{\tt To appear in Physica E, special issue on ``Dynamics of Complex Systems''}\\
\hrule
\end{abstract}

\maketitle

\section{Problem}

The purpose of this note is to answer a question raised by Kogan, Mello, and
Liqun \cite{Kog99}, concerning the statistics of the eigenvalues of the
scattering-matrix product $SS^{\dagger}$ for an absorbing optical cavity with
chaotic dynamics. {\em Without\/} absorption the scattering matrix $S$ is an
$N\mbox{}\times N$ unitary matrix, hence $SS^{\dagger}$ is simply the unit
matrix. {\em With\/} absorption the eigenvalues $R_{1},R_{2},\ldots R_{N}$ of
$SS^{\dagger}$ are real numbers between 0 and 1. What is the probability
distribution $P(\{R_{n}\})$ of these reflection eigenvalues in an ensemble of
chaotic cavities?

In principle, this problem can be solved by starting from the known
distribution of $S$ in the absence of absorption (which is Dyson's circular
ensemble \cite{Bee97}), and incorporating the effects of absorption by a
fictitious lead \cite{Bro97}. What has been calculated in this way is the
distribution $P(\{R_{n}\})$ for small $N$ \cite{Bro97} and the density
$\rho(R)=\langle\sum_{n}\delta(R-R_{n})\rangle$ for large $N$ \cite{Bee99}.
These results have a complicated form, quite unlike those familiar from the
classical ensembles of random-matrix theory \cite{Meh91}. For example, in the
presence of time-reversal symmetry the distribution for $N=2$ is given by
\cite{Bro97}
\begin{eqnarray}
&& P(R_1,R_2)=(1-R_1)^{-4}(1-R_2)^{-4}\nonumber\\
&&\mbox{}\times\exp\bigl[-\gamma(1-R_1)^{-1}
-\gamma(1-R_2)^{-1}\bigr]\nonumber\\
&&\mbox{}\times|R_1-R_2|\,\bigl[\gamma^2(1-{\rm e}^{2\gamma}+\gamma+\gamma {\rm
e}^{2\gamma})\nonumber\\
&&\mbox{}+\gamma(R_1+R_2-2)({\textstyle\frac{3}{2}}-{\textstyle\frac{3}{2}}{\rm
e}^{2\gamma}+2\gamma+\gamma {\rm e}^{2\gamma}+\gamma^2)\nonumber\\
&&\mbox{}+(1-R_1)(1-R_2)(3-3{\rm
e}^{2\gamma}+{\textstyle\frac{9}{2}}\gamma\nonumber\\
&&\mbox{}+{\textstyle\frac{3}{2}}\gamma {\rm
e}^{2\gamma}+3\gamma^2+\gamma^3)\bigr],\label{PNis2}
\end{eqnarray}
where $\gamma$ is the ratio of the mean dwell time $\tau_{\rm d}$ inside the
cavity\footnote{
The mean dwell time is related to the mean frequency interval $\Delta$ of the
cavity modes by $\tau_{\rm d}=2\pi/N\Delta$, so that $\gamma=2\pi(\tau_{\rm
a}N\Delta)^{-1}$. The definition of $\gamma$ used in Ref.\ \protect\cite{Bro97}
differs by a factor $N$.}
and the absorption time $\tau_{\rm a}$.

The situation is simpler for reflection from an absorbing disordered waveguide.
In the limit that the length of the waveguide goes to infinity, the
distribution of the reflection eigenvalues becomes that of the Laguerre
ensemble, after a transformation of variables from $R_{n}$ to
$\lambda_{n}=R_{n}(1-R_{n})^{-1}\geq 0$. The (unnormalized) distribution is
given by \cite{Bee96,Bru96}
\begin{eqnarray}
&&P(\{\lambda_{n}\})\propto\prod_{i<j}|\lambda_{i}-\lambda_{j}|^{\beta}
\nonumber\\
&&\mbox{}\times\prod_{k}\exp[-\gamma(\beta N+2-\beta)\lambda_{k}]
,\label{Pwire}
\end{eqnarray}
where now $\gamma=\tau_{\rm s}/\tau_{\rm a}$ contains the scattering time
$\tau_{\rm s}$ of the disorder. The integer $\beta=1(2)$ in the presence
(absence) of time-reversal symmetry. The eigenvalue density is given by a sum
over Laguerre polynomials, hence the name ``Laguerre ensemble'' \cite{Meh91}.

Kogan, Mello, and Liqun \cite{Kog99} used a maximum entropy assumption
\cite{Mel99} to argue that a chaotic cavity is also described by the Laguerre
ensemble, but in the variables $R_{n}$ instead of the $\lambda_{n}$'s. Their
maximum entropy distribution is
\begin{equation}
P(\{R_{n}\})\propto\prod_{i<j}|R_{i}-R_{j}|^{\beta}
\prod_{k}\exp(-aR_{k}).\label{Pmaxent}
\end{equation}
The coefficient $a$ in the exponent is left undetermined.\footnote{
We have verified that the theory of Ref.\ \protect\cite{Bro97} agrees with Eq.\
(\protect\ref{Pmaxent}) for strong absorption ($\gamma\gg 1$), with coefficient
$a=\frac{1}{2}\gamma\beta N$.}
Comparison with computer simulations gave good agreement for strong absorption,
but not for weak absorption \cite{Kog99}. This is unfortunate since the
weak-absorption regime ($\gamma\ll 1$) is likely to be the most interesting for
optical experiments. Although we know from the exact small-$N$ results
\cite{Bro97} that no simple distribution exists in the entire range of
$\gamma$, one might hope for a simple eigenvalue distribution for small
$\gamma$. What is it?

\section{Solution}

Absorption with rate $1/\tau_{\rm a}$ is equivalent to a shift in frequency
$\omega$ by an imaginary amount $\delta\omega={\rm i}/2\tau_{\rm a}$. If we
denote by $S(\omega)$ the scattering matrix with absorption and by
$S_{0}(\omega)$ the scattering matrix without absorption, then
$S(\omega)=S_{0}(\omega+{\rm i}/2\tau_{\rm a})$. For weak absorption we can
expand
\begin{eqnarray}
S_{0}(\omega+{\rm i}/2\tau_{\rm a})&\approx&S_{0}(\omega)+\frac{\rm
i}{2\tau_{\rm a}}\frac{d}{d\omega}S_{0}(\omega)\nonumber\\
&=&S_{0}(\omega)\left[1-\frac{1}{2\tau_{\rm
a}}Q(\omega)\right],\label{S0Qrelation}
\end{eqnarray}
where $Q=-{\rm i}S_{0}^{\dagger}dS_{0}^{\vphantom\dagger}/d\omega$ is the
time-delay matrix \cite{Fyo97}. Since $S_{0}$ is unitary, $Q$ is Hermitian. The
eigenvalues of $Q$, the delay times $\tau_{1},\tau_{2},\ldots\tau_{N}$, are
real positive numbers. Eq.\ (\ref{S0Qrelation}) implies that, for weak
absorption,
\begin{equation}
S(\omega)S^{\dagger}(\omega)=S_{0}^{\vphantom\dagger}(\omega)
\left[1-\frac{1}{\tau_{\rm a}}Q(\omega)\right]S_{0}^{\dagger}(\omega).\label{SQrelation}
\end{equation}

We conclude that the matrix product $SS^{\dagger}$ is related to the time-delay
matrix $Q$ by a unitary transformation. This relationship is a generalization
to $N>1$ of the result of Ramakrishna and Kumar \cite{Ram99} for $N=1$ (when
the unitary transformation becomes a simple identity). Because a unitary
transformation leaves the eigenvalues unchanged, one has
$R_{n}=1-\tau_{n}/\tau_{\rm a}$, or equivalently, $\lambda_{n}=\tau_{\rm
a}/\tau_{n}$ (since $\lambda_{n}\rightarrow(1-R_{n})^{-1}$ for weak
absorption).

The probability distribution of the delay times in a chaotic cavity has
recently been calculated, first for $N=1$ \cite{Fyo96,Gop96} and later for any
$N$ \cite{Bro99}. The corresponding distribution of the reflection eigenvalues
for weak absorption is a generalized Laguerre ensemble in the variables
$\lambda_{n}$,
\begin{eqnarray}
&&P(\{\lambda_{n}\})\propto\prod_{i<j}|\lambda_{i}-\lambda_{j}|^{\beta}
\nonumber\\
&&\mbox{}\times\prod_{k}\lambda_{k}^{\beta
N/2}\exp[-{\textstyle\frac{1}{2}}\gamma\beta N\lambda_{k}].\label{Pcavity}
\end{eqnarray}
The eigenvalue density is given in terms of {\em generalized\/} Laguerre
polynomials, hence the name. The corresponding distribution of the reflection
eigenvalues is
\begin{eqnarray}
&&P(\{R_{n}\})\propto\prod_{i<j}|R_{i}-R_{j}|^{\beta}\nonumber\\
&&\;\;\mbox{}\times\prod_{k}\frac{\exp[-{\textstyle\frac{1}{2}}\gamma\beta
N(1-R_{k})^{-1}]}{(1-R_{k})^{2-\beta+3\beta N/2}}.\label{PcavityR}
\end{eqnarray}
The first moment of this distribution is
$N^{-1}\langle\sum_{n}R_{n}\rangle=1-\gamma$, independent of $\beta$. One can
check that Eq.\ (\ref{PcavityR}) is the small-$\gamma$ asymptote of the exact
result (\ref{PNis2}) for $N=2$, $\beta=1$.

In the case $N=1$ of a single scattering channel the distribution
(\ref{PcavityR}) reduces to
\begin{eqnarray}
&&P(R)=\frac{(\gamma\beta/2)^{1+\beta/2}}{\Gamma(1+\beta/2)}(1-R)^{-2-\beta/2}
\nonumber\\
&&\mbox{}\times\exp[-{\textstyle\frac{1}{2}}\gamma\beta
(1-R)^{-1}],\label{PcavityNis1}
\end{eqnarray}
including the normalization constant. We have plotted this function in Fig.\ 1
for $\gamma=0.1$ and $\beta=1,2$. It is totally different from the exponential
distribution $P(R)\propto\exp(-aR)$ of Ref.\ \cite{Kog99}. For comparison, we
have also included in Fig.\ 1 the exact $N=1$ result \cite{Bro96} (which is
known only for $\beta=2$):
\begin{eqnarray}
&&P(R)=(1-R)^{-3}\exp[-\gamma(1-R)^{-1}]\nonumber\\
&&\mbox{}\times\left[\gamma({\rm e}^{\gamma}-1)+(1+\gamma-{\rm
e}^{\gamma})(1-R)\right].\label{PNis1}
\end{eqnarray}
It is indeed close to the small-$\gamma$ asymptote (\ref{PcavityNis1}).

\begin{figure}[ht]
\centerline{
\psfig{figure=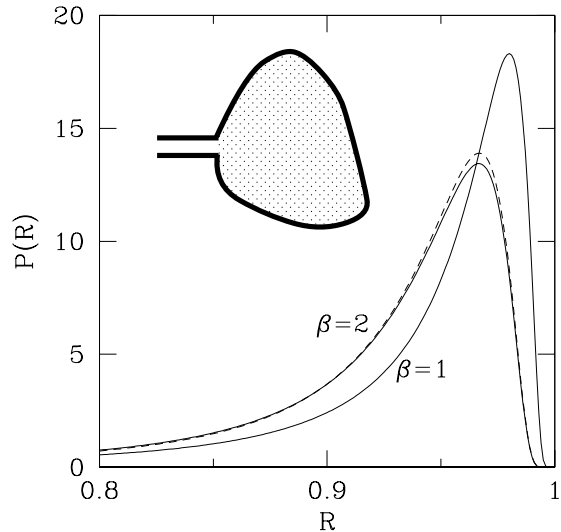,width= 8cm}}
\medskip
\caption[]{
Probability distribution of the reflectance of a weakly absorbing chaotic
cavity that is coupled to the outside via a single-mode waveguide (inset). The
solid curves have been computed from Eq.\ (\protect\ref{PcavityNis1}) for
$\gamma=0.1$ (hence $\langle R\rangle=1-\gamma=0.9$). The dashed curve is the
exact $\beta=2$ result from Eq.\ (\protect\ref{PNis1}).
}
\end{figure}

\section{Conclusion}

Summarizing, the distribution of the reflection eigenvalues of a weakly
absorbing chaotic cavity is the generalized Laguerre ensemble (\ref{Pcavity})
in the parametrization $\lambda_{n}=R_{n}(1-R_{n})^{-1}$. The Laguerre ensemble
(\ref{Pmaxent}) in the variables $R_{n}$, following from the maximum entropy
assumption \cite{Kog99}, is only valid for strong absorption. For intermediate
absorption strengths the distribution is not of the form of the Laguerre
ensemble in any parametrization, cf.\ Eq.\ (\ref{PNis2}). In contrast, the
distribution of a long disordered waveguide is the Laguerre ensemble
(\ref{Pwire}) for all absorption strengths.

The relationship between the reflection eigenvalues for weak absorption and the
delay times implies that the delay times $\tau_{n}$ for reflection from a
disordered waveguide of infinite length are distributed according to Eq.\
(\ref{Pwire}) if one substitutes $\gamma\lambda_{n}\rightarrow\tau_{\rm
s}/\tau_{n}$. The implications of this Laguerre ensemble for the delay times
will be discussed elsewhere.

\section*{Acknowledgements}

Correspondence with E. Kogan and P.A. Mello has stimulated us to look into this
problem. This research was supported by the ``Ne\-der\-land\-se
or\-ga\-ni\-sa\-tie voor We\-ten\-schap\-pe\-lijk On\-der\-zoek'' (NWO), by the
``Stich\-ting voor Fun\-da\-men\-teel On\-der\-zoek der Ma\-te\-rie'' (FOM),
and by the National Science Foundation (grant numbers DMR 94-16910, DMR
96-30064, DMR 97-14725).

\end{document}